\documentclass[ajl]{emulateapj}

\usepackage{graphicx,epsfig,natbib,color}
\usepackage{lscape}
\usepackage{graphicx}
\usepackage{epsfig}
\usepackage{natbib}

\definecolor{DarkGreen}{rgb}{0.0,0.45,0.0}

\newcommand{\apjsub} {{ApJ submitted}}

\slugcomment{Accepted for publication in ApJ}

\begin{document}
\title{Formation of a Double-decker Magnetic Flux Rope in the Sigmoidal Solar Active Region 11520}

\author{X. Cheng$^{1,2,3}$, M. D. Ding$^{1,3}$, J. Zhang$^{1,4}$, X. D. Sun$^{5}$, Y. Guo$^{1,3}$, Y. M. Wang$^{6}$, B. Kliem$^{7,8}$, \& Y. Y. Deng$^{2}$}

\affil{$^1$ School of Astronomy and Space Science, Nanjing University, Nanjing 210093, China}\email{xincheng@nju.edu.cn}
\affil{$^2$ Key Laboratory of Solar Activity, National Astronomical Observatories, Chinese Academy of Sciences, Beijing 100012, China}
\affil{$^3$ Key Laboratory for Modern Astronomy and Astrophysics (Nanjing University), Ministry of Education, Nanjing 210093, China}
\affil{$^4$ School of Physics, Astronomy and Computational Sciences, George Mason University, Fairfax, VA 22030, USA}
\affil{$^5$ W. W. Hansen Experimental Physics Laboratory, Stanford University, Stanford, CA 94305, USA}
\affil{$^6$ School of Earth and Space Sciences, University of Science and Technology of China, Hefei 230026, China}
\affil{$^7$ Institute of Physics and Astronomy, University of Potsdam, D-14476 Potsdam, Germany}
\affil{$^8$ Yunnan Observatory, Chinese Academy of Sciences, P. O. Box 110, Kunming, Yunnan 650011, China}

\begin{abstract}
In this paper, we address the formation of a magnetic flux rope (MFR) that erupted on 2012 July 12 and caused a strong geomagnetic storm event on July 15. Through analyzing the long-term evolution of the associated active region observed by the Atmospheric Imaging Assembly and the Helioseismic and Magnetic Imager on board the \textit{Solar Dynamics Observatory}, it is found that the twisted field of an MFR, indicated by a continuous S-shaped sigmoid, is built up from two groups of sheared arcades near the main polarity inversion line half day before the eruption. The temperature within the twisted field and sheared arcades is higher than that of the ambient volume, suggesting that magnetic reconnection most likely works there. The driver behind the reconnection is attributed to shearing and converging motions at magnetic footpoints with velocities in the range of 0.1--0.6 km s$^{-1}$. The rotation of the preceding sunspot also contributes to the MFR buildup. Extrapolated three-dimensional non-linear force-free field structures further reveal the locations of the reconnection to be in a bald-patch region and in a hyperbolic flux tube. About two hours before the eruption, indications for a second MFR in the form of an S-shaped hot channel are seen. It lies above the original MFR that continuously exists and includes a filament. The whole structure thus makes up a stable double-decker MFR system for hours prior to the eruption. Eventually, after entering the domain of instability, the high-lying MFR impulsively erupts to generate a fast coronal mass ejection and X-class flare; while the low-lying MFR remains behind and continuously maintains the sigmoidicity of the active region.
\end{abstract}

Online-only material: animations, color figures

\section{Introduction}
A magnetic flux rope (MFR) is defined as a current channel with a set of magnetic field lines wrapping more than once around the central axis. Such a structure often erupts from the Sun as a coronal mass ejection (CME; the largest-scale eruption in the solar system), which releases a large quantity of magnetized plasma into the interplanetary space with a velocity of hundreds km s$^{-1}$, even up to 3000 km s$^{-1}$ \citep{yashiro04,zhang06,chen11_review}. The magnetized plasma is still organized frequently as a coherent MFR when arriving at the Earth, as indicated by features such as the magnetic field rotation, the drop of the solar wind density and proton temperature, and the low plasma beta in the in situ observations \citep{burlaga81}.

Several lines of evidence imply that the MFR exists in the corona prior to the CME eruption. Sigmoid, a forward or reversed S-shaped emission pattern in soft X-ray (SXR) and extreme ultraviolet (EUV) passbands, often appears in CME-productive active regions (ARs) \citep{rust96,canfield99}. The straight section in the middle of the sigmoid is believed to be strong evidence of the MFR existing in the corona \citep[e.g.,][]{sterling00,liuc07,mckenzie08,liur10,savcheva12a}. Filaments are another piece of evidence of the existence of the MFR, which includes magnetic dips that are able to collect cool material against gravity \citep{mackay10,guo10_filament,suyingna11,suyingna12}. Filament channels are even thought to be the body of the MFR because rotation motions are often observed at the bottom of dark cavities, the cross sections of filament channels at the solar limb \citep{low95_apj,guo98,gibson04,wangym10,lixing12}. Moreover, taking advantage of the observed photospheric vector magnetic field at the bottom boundary, the MFR can be reconstructed by extrapolation techniques using the assumption of a non-linear force free field (NLFFF); this has also indicated preexistence of the MFR \citep[e.g.,][]{yan01,canou09,guo10_filament,cheng10_reflare,cheng13_double,suyingna11,jiang13,jiang14_nlfff,inoue13}.

If the MFR really exists in the corona prior to eruption, the question arises of when and where the MFR is built up. Two possibilities are proposed theoretically. One is that the MFR is generated in the convection zone and partly emerges into the corona by buoyancy \citep{fan01,Martnez-sykora08}. \citet{manchester04} noted that when the primary axis approaches the photosphere, the MFR may split into two parts by the reconnection with the surrounding fields, which only allows the upper part of the MFR to ascend to the corona \citep[also see][]{magara06,archontis08_aa,leake13}. The other possibility is that the MFR is built up directly in the corona. Through imposing shearing and/or vortex motions to the different polarities, an initial potential field is gradually sheared. Converging flows then initiate the reconnection near the polarity inversion line (PIL), which converts sheared fields into twisted ones \citep{vanballegooijen89,amari03a}. Depending on the specific locations of the reconnection, the MFR is created either through flux cancellation in the photosphere prior to the eruption \citep{aulanier10,amari11,xia14} or via tether-cutting and flare reconnection in the corona during the eruption \citep{moore01,antiochos99,lynch08,karpen12}.

Observationally, the MFR is also conjectured either to stem from emergence from below the photosphere or to build up in the corona. \citet{lites05} studied the properties of the vector magnetograms associated with two filaments and found a concave up geometry underneath the filaments. \citet{okamoto08} examined a sequence of vector magnetograms of AR 10953 and found that two abutting opposite-polarity regions with horizontally strong but vertically weak magnetic fields grew laterally and then narrowed. The directions of the horizontal magnetic fields along the PIL gradually reversed from a normal polarity to an inverse one. Both concave up geometry and reversed polarity suggest that the MFR may come from below the photosphere. However, \citet{Dominguez12} recently provided an opposite interpretation for the photospheric evolution characteristics of the AR 10953. Through comparing with the numerical results by \citet{mactaggart10}, they stated that magnetic cancellation is also able to produce the lateral growing and then narrowing of the opposite polarities, as well as the reversal of the horizontal field direction. Moreover, through analyzing the temperature structure of a sigmoid, \citet{tripathi09} discovered that the plasma in the J-shaped arcades can have a higher temperature than that in the S-shaped flux if both are simultaneously visible. They argued that it is most likely that the J-shaped arcades are reconnecting to the S-shaped flux. The reconnection at the same time heats the plasma, which afterwards enters a cooling phase. \citet{green09} and \citet{green11} supported the conjecture that the reconnection is mostly associated with flux cancellation, although only part of the cancelled flux may be injected into the MFR. It is worth noting that the MFR can even be formed during a confined flare and be destabilized in a subsequent major eruption \citep{patsourakos13,song14}. This is similar to the conjecture of \citet{guo13_qmap}, who argued that the quasi-separatrix layer reconnection in the interface between the MFR and the surrounding fields, indicated by a series of confined flares, has an important role in injecting self-helicity to the MFR.

Although previous works have displayed elementary results on the formation of the MFR, their objects for study are filaments or sigmoids, not the MFR itself. Recently, using the data from the Atmospheric Imaging Assembly \citep[AIA;][]{lemen12} on board the \textit{Solar Dynamics Observatory} (\textit{SDO}), \citet{zhang12} and \citet{cheng13_driver} found that the MFR directly exists as an elongated EUV channel structure appearing in the high temperature AIA passbands at 131 and 94 {\AA}. In the impulsive acceleration phase, the MFR is further enhanced by flare reconnection, and the morphology evolves from the sigmoidal shape to the semicircular one. A number of further observations have indicated that this hot channel is actually the MFR that plays an important role in forming and accelerating the CME \citep{zhang12,cheng13_driver,cheng14_tracking,lileping13}. In this paper, we pay our attention to the long-term formation process of a hot channel in the sigmoidal AR NOAA 11520. We find that (1) the MFR is most likely formed through coexisting reconnection at a bald patch (BP) and at a hyperbolic flux tube, driven by photospheric shearing and converging flows; (2) a second high-lying MFR (the hot channel) locates vertically above a filament-associated low-lying MFR; the whole structure thus constitutes a double-decker MFR system that stably exists for about two hours prior to the eruption. The instruments are presented in Section 2. Observations of the MFR formation are displayed in Section 3, followed by the causes of the MFR formation in Section 4. In Section 5, we give our summary and discussions.

\section{Instruments}
The data used in this study are mainly from the AIA \citep{lemen12} and the Helioseismic and Magnetic Imager \citep[HMI;][]{schou12}, both of which are on board the \textit{SDO}. The AIA includes ten passbands, six of which image the solar corona almost simultaneously with an unprecedented high cadence (12 seconds) and high spatial resolution (1.2{\arcsec}), covering the temperature range from 0.06 MK to 20 MK \citep{odwyer10}. The HMI observes the vector magnetic field of the full solar photosphere with approximately the same spatial resolution (1.0{\arcsec}) as the AIA but with a cadence of 12 minutes. The X-Ray Telescope \citep[XRT;][]{golub07} aboard \textit{Hinode} \citep{kosugi07} images the hot plasma in the corona. The \textit{Geostationary Operational Environmental Satellite} (\textit{GOES}) and the \textit{Reuven Ramaty High Energy Solar Spectroscopic Imager} \citep[\textit{RHESSI};][]{linrp02} spacecraft register the SXR and hard X-ray (HXR) fluxes of solar flares. In addition, the Large Angle and Spectrometric Coronagraph \citep[LASCO;][]{brueckner95} on board the \textit{Solar and Heliospheric Observatory} (\textit{SOHO}) and the Sun-Earth Connection Coronal and Heliospheric Investigation \citep[SECCHI;][]{howard08} on board the \textit{Solar Terrestrial Relations Observatory} (\textit{STEREO-A} and \textit{STEREO-B}) provide the EUV and white-light images of the CME.

\begin{figure*}
\center {\includegraphics[width=15cm]{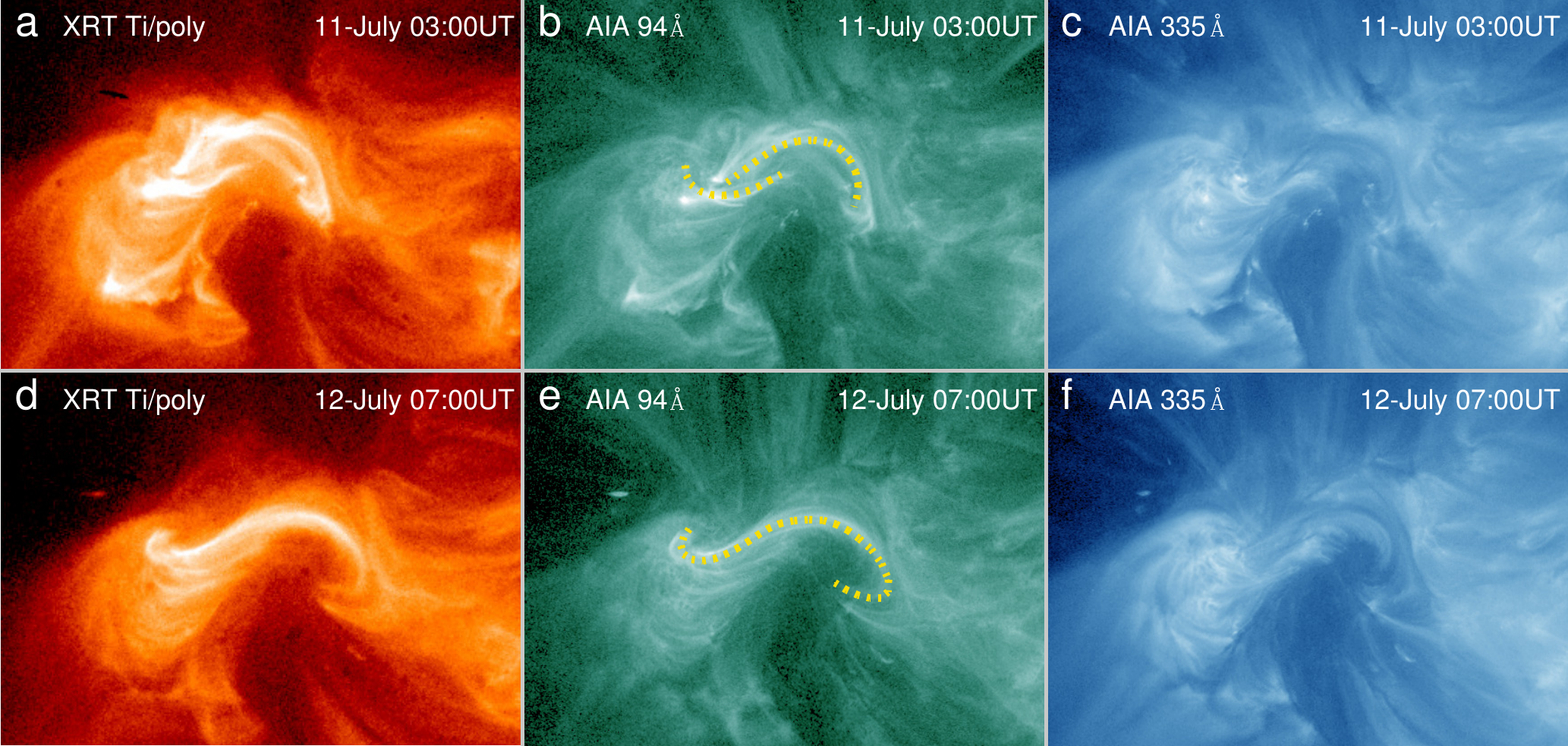}}
\caption{\textit{Hinode}/XRT Ti-poly (a and d) and \textit{SDO}/AIA 94 {\AA} (b and e) and 335 {\AA} images (c and f) showing the formation of the MFR in the sigmoidal AR 11520. The two J-shaped yellow dotted lines and the S-shaped one depict two bundles of strong sheared arcades and the twisted field, respectively.}
(Animations this figure are available in the online journal.)
\label{f1}
\end{figure*}

\begin{table*}
\caption{Phases of the formation and development of the double-decker MFR.}\vspace{0.0\textwidth}
\label{tb1}
\begin{tabular}{l p{10cm}}\tableline \tableline
 Time        &  Note  \\
\hline
July 7  00:00 UT        & The AR 11520 rotated from backside to the east limb. \\
July 11 00:00 UT       & A sigmoidal structure appeared, the core field mainly consists of two sets of hot sheared arcades (Figure \ref{f1}a--\ref{f1}c, Figure \ref{f6}g, and Figure \ref{f8}a). \\
July 11 00:00 -- July 12 03:00 UT       & The sheared arcades transformed to the continuous S-shaped field lines (Figure \ref{f1}d--\ref{f1}f, Figure \ref{f6}h, and Figure \ref{f8}b) driven by the shear and convergence flows near the main neutral line (Figure \ref{f7}). \\
July 12 14:00 UT       & A diffuse and elongated high-lying channel appeared (Figure \ref{f2}a--\ref{f2}c and Figure \ref{f3}). \\
July 12 16:10 UT       & The hot channel started to expand and rise (Figure \ref{f2}g--\ref{f2}i), resulting in a CME (Figure \ref{f4}) and X1.4 class flare, but the filament survived (Figure \ref{f5}c).\\
July 13 14:00 UT       & The low-lying MFR and associated filament could be identified with the disappearance of flare loops (Figure \ref{f5}f and Figure \ref{f8}d). \\
July 15 06:00 UT       & The high-lying MFR-associated magnetic cloud arrived at the Earth. \\
July 15 19:00 UT       & The Dst index peaks at $-127$~nT
.   \\
\tableline
\vspace{0.01\textwidth}
\end{tabular}
\end{table*}

\section{Observations of the MFR}

\subsection{Formation of double-decker MFR}
\label{3.1}
On 2012 July 15, a magnetic cloud reached the Earth at $\sim$06:00 UT and then gave rise to a strong geomagnetic storm event with the minimal Dst index of --127 nT ($\sim$19:00 UT). Inspecting the AIA data carefully, we find that the magnetic cloud was related to a strong solar eruption that occurred on July 12 and produced an X1.4 class flare and a fast halo CME. The flare was located at the heliographic coordinates S17W08. The corresponding source region is NOAA active region (AR) 11520. The flare SXR flux started to rise gradually at $\sim$14:50 UT, then rapidly increased from $\sim$16:10 UT and peaked at $\sim$16:49 UT.

In order to investigate the origin of the eruption, we examine the AIA images of the AR in all EUV passbands from July 6 to July 13. The overall evolution of the event is summarized in Table \ref{tb1}. It is found that the AR appeared at the solar limb on July 7 and progressively displayed a sigmoidal structure from July 11 onward. From the XRT Ti-poly and the AIA 94 {\AA} images (Figure \ref{f1}a--\ref{f1}b), one can see that initially the sigmoid was mainly composed of two groups of J-shaped arcades, whose straight arms were located at the two sides of the PIL with the elbows crossing the PIL at the opposite ends. From $\sim$03:00 UT on 2012 July 12, the two groups of J-shaped arcades gradually transformed to continuous S-shaped field lines (Figure \ref{f1}d--\ref{f1}e), which is a strong indication that an MFR has formed \cite[e.g.,][]{savcheva12b}. In the AIA 335 {\AA} images, we only see an overall outer envelope of the sigmoid, in which the detailed evolution of the coronal structure is difficult to detect (Figure \ref{f1}c and \ref{f1}f). These results show that the formation of the MFR by reconnection of J-shaped arcades mainly occurs in the core field of the sigmoid, where the plasma has been heated to the high temperatures ($\ge$10 MK); while for the envelope of the sigmoid, the plasma temperature is relatively lower ($\sim$2.5 MK), thus not indicating any strong reconnection. Moreover, in the AIA 304 {\AA} passband, we notice that a J-shaped filament was visible from June 6 onward, located in the right part of the sigmoid. Its presence also indicates high shear or even twist of the sigmoid.  

\begin{figure*}
\center {\includegraphics[width=15cm]{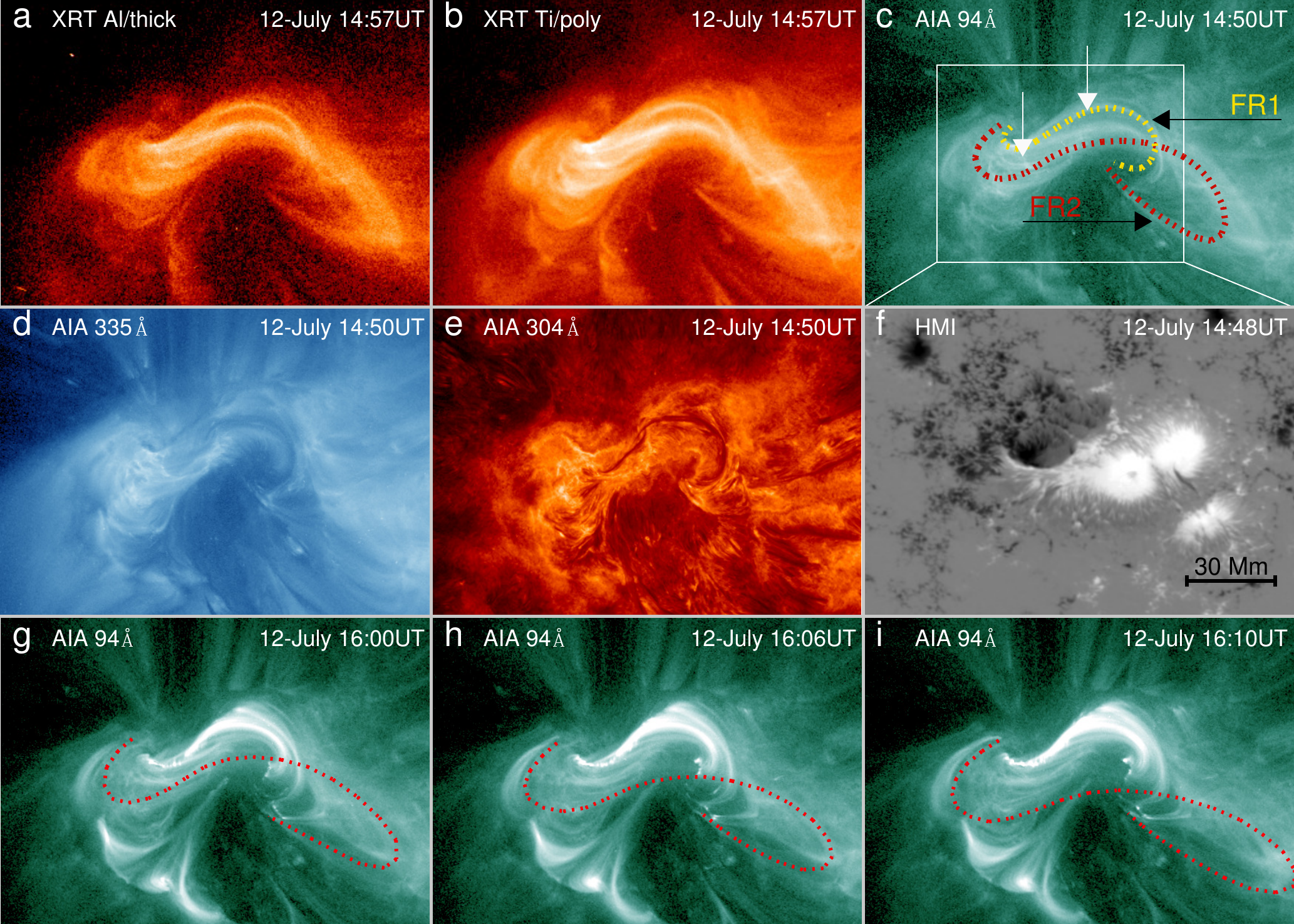}\vspace{-0.0\textwidth}}
\caption{\textit{Hinode}/XRT and \textit{SDO}/AIA images displaying the double-decker MFR configuration before the eruption. a--f: XRT Al-thick, Ti-poly, the AIA 94 {\AA}, 335 {\AA}, 304 {\AA}, and the HMI line-of-sight magnetogram. The two S-shaped dotted lines indicate the low-lying (yellow) MFR and the high-lying MFR (red), respectively. The field-of-view of the panel f is pointed out by the box in the panel c. g--i: AIA 94 {\AA} images showing the expansion and rising of the over-lying MFR in the early eruption phase, as indicated by the red dotted lines.}
(Animations this figure are available in the online journal.)
\label{f2}
\end{figure*}

\begin{figure*}
\center {\includegraphics[width=15.3cm]{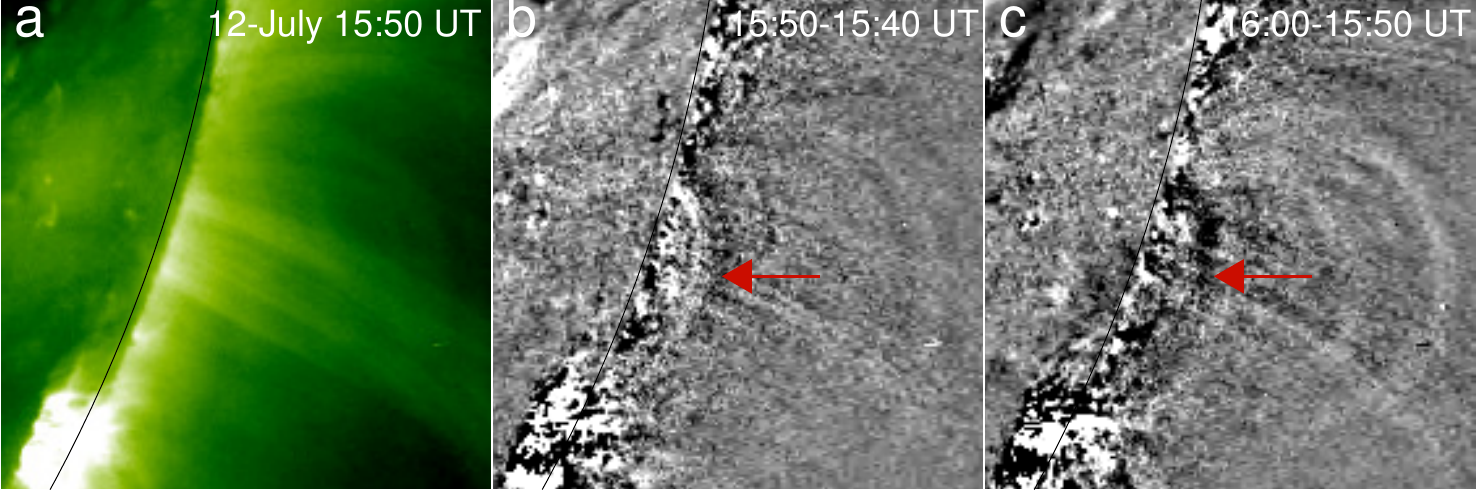}}\\
\caption{a: \textit{STEREO}/EUVI-B 195 {\AA} image from which no filament is seen. b and c: \textit{STEREO}/EUVI-B 195 {\AA} running-difference images. The red arrows show the initial height of the high-lying MFR. The black lines display the optical solar limb.}
(Animation this figure is available in the online journal.)
\label{f3}
\end{figure*}

A closer inspection of the data leads us to the conclusion that a high-lying MFR and a low-lying MFR coexisted above the same PIL of the AR for two hours prior to the eruption. From $\sim$14:00 UT on July 12, we can see a diffuse and hot EUV channel structure in the AIA 94 {\AA} passband (elongated S-shaped; outlined by the red dotted line in Figure \ref{f2}c). The temperature of the channel is higher than 6 MK because it can only be seen in the hot temperature passbands (i.e., AIA 94 and 131 {\AA} and XRT Ti-poly) but not in the other lower temperature ones (Figure \ref{f2}d). From $\sim$16:00~UT, the hot channel showed an accelerating expansion (Figure~\ref{f2}g--\ref{f2}i), with the elbows expanding eastward and westward and the middle moving to the south. Because the channel is diffuse and less bright than the flare signatures, only the animation of the AIA 94~{\AA} images that accompanies Figure~\ref{f2} permits a full appreciation of the hot channel's shape and dynamics. While the hot channel erupted, the filament (as seen in Figure \ref{f2}e) stayed in place; it did not show any displacement larger than the slight changes in position seen during the preceding days. Such partial sigmoid eruptions are not uncommon \citep{pevtsov02} and have been interpreted as a partial eruption of an MFR, whose top part has become unstable while the bottom part is tied to the photosphere in a BP \citep{gilbert01,gibson06_apjl,gibson08_jgr,green09}. While this is a plausible scenario in general, here it runs into difficulties because the HMI vector field data presented below show a BP section of the PIL only under the shorter left part of the filament (see, e.g., Figure~\ref{f9}a below). Moreover, at least some motion of the filament was likely to occur if it was part of the same erupting MFR as the hot channel; however, no motion in the southward direction of the eruption was seen. Next we consider the arcade of loops that dominated the middle and left part of the sigmoid in the AIA 94~{\AA} images prior to the eruption of the hot channel (marked by white arrows in Figure~\ref{f2}c). These loops did not show any systematic or significant change in position or shape while the hot channel erupted (see Figure~\ref{f2}g--\ref{f2}i and the accompanying animation). Such a behavior is clearly at variance with the interpretation that a single MFR erupted in part. Finally, we consider the perspective of \textit{STEREO-B} (Figure \ref{f3}a) and find that the filament was blocked behind the solar limb, indicating it is low-lying; while the counterpart of the hot channel in the EUV 195 {\AA} passband, probably from the emission of the \ion{Fe}{24} line at 192 {\AA} \citep{milligan13}, is located at a height of $\sim$90 Mm above the limb in the period of 14:00--16:00 UT, which is far above typical heights of active-region filaments \citep[e.g.,][]{tandberg-hanssen95}. Thus, we are led to conclude that the hot channel was independent of the magnetic flux of the filament-associated low-lying MFR and constituted a double-decker MFR system already prior to the onset of the impulsive phase of the eruption. Note that the identification of the structure of the filament with a low-lying MFR is further supported by the NLFFF modeling in Section \ref{ss:NLFFF}.

\subsection{Partial Eruption of double-decker MFR}

\begin{figure*}
\center {\includegraphics[width=15cm]{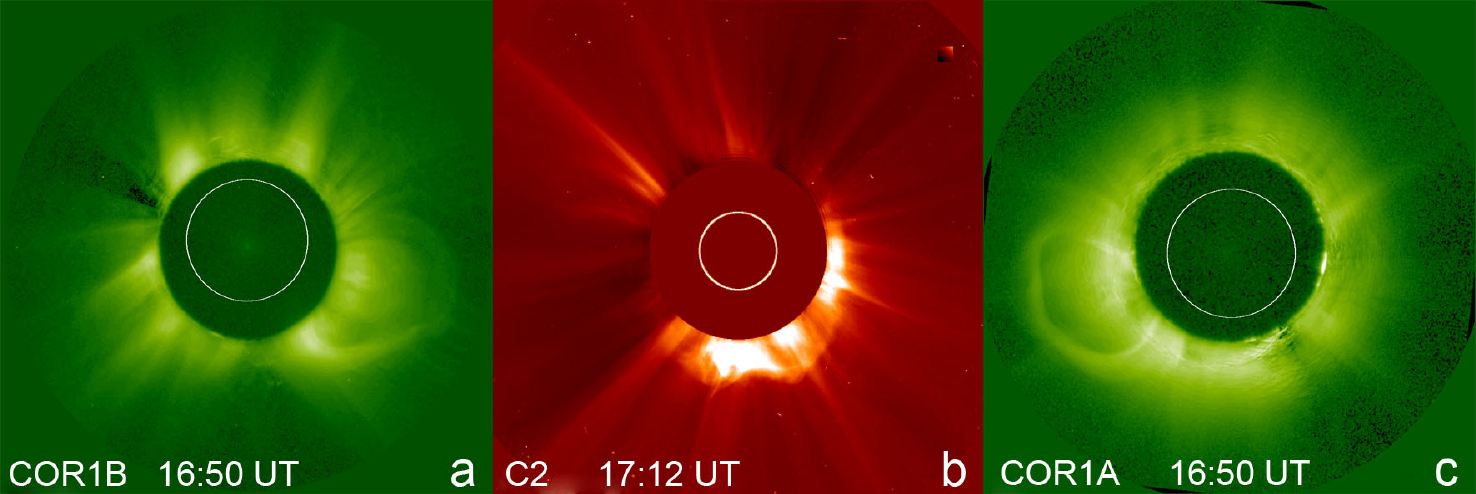}}\\
\caption{\textit{STEREO}/COR1 and \textit{SOHO}/C2 white-light images of the CME on 2012 July 12. The white circles indicate the solar limb.}
\label{f4}
\end{figure*}

\begin{figure*}
\vspace{-0.0\textwidth}
\center {\includegraphics[width=15cm]{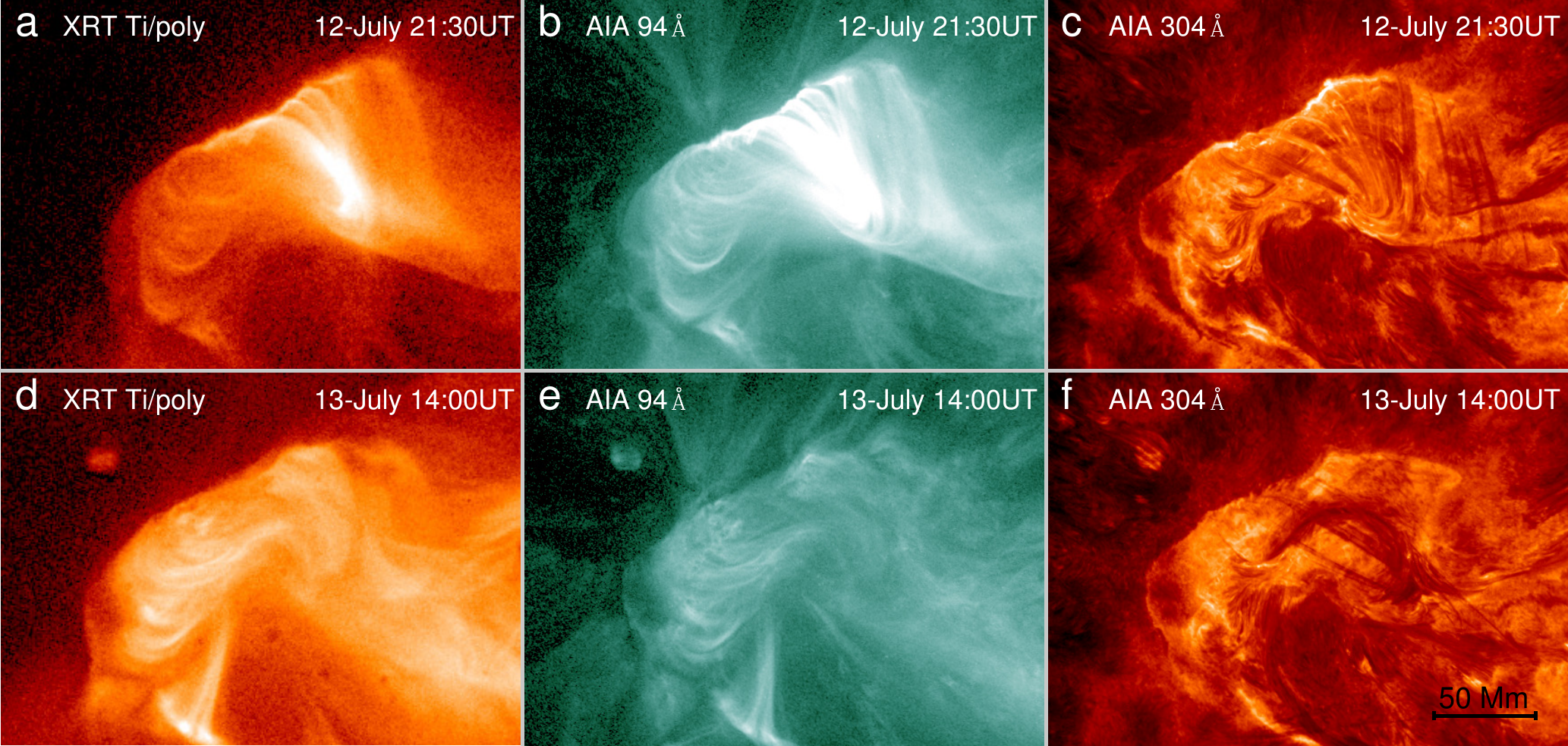}\vspace{-0.0\textwidth}}\\
\caption{\textit{Hinode}/XRT Ti-poly and \textit{SDO}/AIA 94 {\AA} and 304 {\AA} images showing the post-flare loops (a--c) and the reappearance of the sigmoid after the eruption (d--f).}
\label{f5}
\end{figure*}

The explosive eruption of the high-lying MFR commenced at $\sim$16:10 UT, which resulted in the quick formation and acceleration of a CME and the rapid enhancement of the flare emission at various wavelengths. The CME first appeared as a limb event in the field-of-view (FOV) of the SECCHI coronagraphs COR1-A and COR1-B at $\sim$16:25 UT, and as a halo CME in the FOV of LASCO/C2 at $\sim$17:00 UT (Figure \ref{f4}). As for the flare, we note that the most remarkable feature from $\sim$15:00 UT to 16:10 UT is the brightening of the right arcade of the sigmoid and of the footpoints of the double-decker MFR, which are closely related to the slow rise and accelerating expansion of the high-lying MFR (Figure \ref{f2}g--\ref{f2}i and attached high cadence movies). After 16:10 UT, the footpoints brightenings extended along the direction of the PIL. At $\sim$16:25 UT, the brightenings formed two ribbons in the chromosphere, which separated from each other afterwards as seen in the movies of the AIA 304, 1600, and 1700 {\AA} passbands (see details in attached high cadence movies of Figure \ref{f2}). From the AIA 94 {\AA} and XRT Ti-poly images, we can see the arcade of flare loops, the ends of which corresponded well to the two separating ribbons (Figure \ref{f5}a--\ref{f5}b). Underneath the flare loops, the whole filament stayed quietly with the inferred low-lying MFR regardless of the eruption of the high-lying MFR. Moreover, we find that the CME in the COR1 images (Figure 4) mainly consisted of a bright front and a relatively diffuse cavity. We cannot, however, detect a clear appearance of a bright core that is generally thought to be erupted filament material and located at the bottom part of the cavity \citep{illing83,vourlidas13}. This seems to also support that the filament did not erupt, or at least did not fully erupt. Mostly, the whole filament structure was not influenced much by the high-lying MFR eruption.

With the cooling and disappearance of the flare loops, the signatures of the low-lying MFR again showed up and redisplayed the sigmoidicity of the AR on July 13. In the high temperature passbands (AIA 94 {\AA} and XRT Ti-poly; Figure \ref{f5}d and \ref{f5}e), one can obviously see that some S-shaped field lines became visible in the AR. The survived filament was located in the middle of the sigmoid as seen in the AIA 304 {\AA} passband (Figure \ref{f5}f).

\section{Origin of the MFR}

\subsection{DEM Properties of MFR}
Thanks to the multi-passband multi-temperature imaging ability of AIA, differential emission measure (DEM) structures of the plasma in the AR can be constructed. The observed flux $F_{i}$ for each AIA passband is given by $F_{i}$=$\int R_{i}(T)\times \rm DEM(\textit{T})\,$d$T$, where $R_{i}(T)$ denotes the temperature response function of passband $i$. In order to reduce the error of the DEM inversion, we first calibrate six near-simultaneous AIA EUV images to the data level 1.5 using the SolarSoft routine ``aia{\_}prep.pro" and then degrade the resolution to 2.4$\arcsec$ by the routine ``rebin.pro", thus guaranteeing a good coalignment accuracy of 0.6{\arcsec} between the images in different passbands \citep{aschwanden13_solarphy}. Using the routine ``xrt{\_}dem{\_}iterative2.pro" as proposed by \citet{weber04} and \citet{golub04}, we reconstruct the DEM in each pixel. The validation of this inversion method and uncertainties of the DEM results can be found in \citet{cheng12_dem}. 

\begin{figure*}
\vspace{-0.0\textwidth}
\center {\includegraphics[width=15cm]{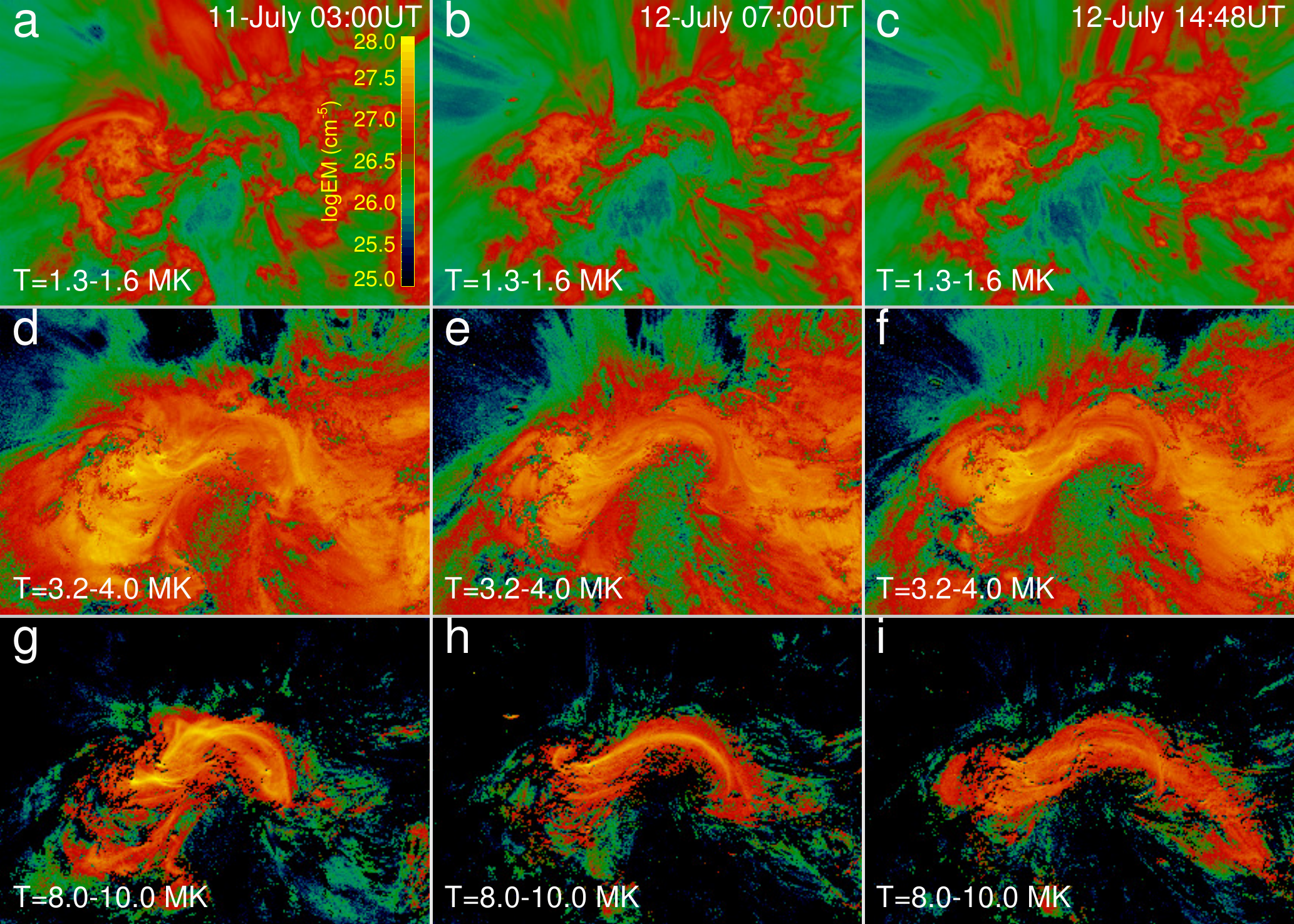}\vspace{-0.0\textwidth}}\\
\caption{EM maps of the sigmoidal AR at different temperature intervals and instants. Panels a--c, d--f, and g--i correspond to the integrated temperature range of 1.3--1.6 MK, 3.2--4.0 MK, and 8.0--10.0 MK, respectively.}
\label{f6}
\end{figure*}

With the DEM in each pixel, we calculate the emission measure (EM) at the different temperature intervals ($\Delta T$) through the formula EM($T$)=$\int_{T-\Delta T}^{T}{\rm DEM}(T^\prime)$d$T^\prime$ to construct the two-dimensional maps of the plasma EM. Figure \ref{f6} shows the EM structure of the AR in three temperature ranges. One can see that the emission in the ambient volume of the AR is dominated by cool plasma (1--2 MK; upper row), whereas the emission of the sigmoidal structure is mostly from warm plasma (3--4 MK; middle row). Hot plasma (8--10 MK) contributes the significant emission in the sigmoid center (bottom row).

The EM maps of the AR provide important clues for understanding the formation of the MFR. From the EM maps in the early phase of the sigmoid, e.g., at 03:00 UT on July 11 (Figure \ref{f6}d), one can only see an indication of the sigmoidal emission pattern at the warm temperature; while at the hot temperature, two groups of clearly sheared sigmoidal arcades have already appeared. Their maximum EM is $\sim$10$^{28}$ cm$^{-5}$, being comparable to that of the loops in the warm temperature range. With the time elapsing, a clear sigmoidal emission pattern appeared at 07:00 UT on July 12 in the warm and hot temperature ranges, possibly due to the fact that magnetic flux carrying heated plasma was added to the continuous S-shaped field (Figure \ref{f6}h). This demonstrates the gradual pre-eruption evolution of the AR, forming an MFR by reconnection. The low-lying MFR is very well captured in the EM maps, and a trace of the high-lying MFR can be identified as well in Figure~\ref{f6}i by the correspondence with the diffuse S-shaped structure marked by the red line in the AIA 94~{\AA} image in Figure~\ref{f2}c. Finally, we note that the peak EM of the low-lying MFR at 10~MK decreased in Figure~\ref{f6}i (to $\sim$10$^{27}$ cm$^{-5}$); this may correspond to the temporarily reduced rate of flux cancellation (Figure \ref{f7}a).

\subsection{Shearing and Converging Flows and Sunspot Rotation}\label{ss:HMI}
In this section, we study the photospheric properties during the MFR formation. We first plot the evolution of the line-of-sight magnetic flux of the sigmoidal AR in Figure \ref{f7}a. During 40 hours before the eruption, the positive flux increased from $\sim$3.6$\times$10$^{22}$ Mx to $\sim$4.0$\times$10$^{22}$ Mx, while the negative flux decreased slightly from $\sim$1.7$\times$10$^{22}$ Mx to $\sim$1.5$\times$10$^{22}$ Mx. Overall, there is no significant flux emergence in the period of the MFR formation (Table \ref{tb1} and Figure \ref{f7}a). After the eruption, the positive flux approximately kept a constant value of $\sim$4.0$\times$10$^{22}$ Mx, whereas the negative flux quickly decreased to $\sim$1.0$\times$10$^{22}$ Mx at 00:00 UT on July 14. The decrease is most likely to be attributed to magnetic cancellation along the main PIL as shown in Figure \ref{f7}f--\ref{f7}i.

In order to investigate the driver of the magnetic cancellation, we calculate the transverse velocity of flux elements in the photosphere through applying the differential affine velocity estimator (DAVE) to a temporal sequence of HMI line-of-sight magnetograms. This technique assumes an affine velocity profile within a small window and then executes variational principles to minimize the deviations in the magnitude of the magnetic induction equation \citep{schuck05,schuck06}. The only free parameter is the window size, which is set to 10 pixels in our calculation. According to the tests by \citet{chae08}, this window size ensures a relative error less than 20\%. 

We find that the MFR formation in the sigmoid is associated with three main types of photospheric motions: shearing, converging, and rotational flows (see Figure \ref{f7}f--\ref{f7}i and the accompanying movie). In the center of the sigmoid there are strong flows toward the PIL, mainly from the positive-polarity side. These consist of the moat flow of the preceding sunspot and of the motion of the sunspot as a whole toward the PIL, both of which persist throughout the time range analyzed. These flows lead to strong magnetic cancellation in the middle of the sigmoid, between the straight legs of the initial double-J pair of arcades. Under the left part of the sigmoid, positive flux is intruding along the PIL toward the center of the AR with a velocity of $\sim$0.3 km s$^{-1}$; while the adjacent negative flux shows motions in the opposite direction at a velocity of $\sim$0.2 km s$^{-1}$ during part of the time. This leads to strong shear flows at a section of the PIL that likely cause significant cancellation, since the polarities are closely butted against each other while possessing a strong relative motion. The shearing motion stretched the loops to form the two sets of J-shaped arcades (Figure \ref{f7}b and \ref{f7}f). The converging flows initiated the reconnection between their heads and tails, producing the twisted field lines as suggested by \citet{martens01}. With the continuous photosphere-driven reconnection, more and more J-shaped arcades are converted to the twisted flux and came into being the MFR (Figure \ref{f7}c and \ref{f7}g). 

In addition to the shearing and converging motions, the rotation of the sunspot also has a significant role in building up the MFR. From Figure \ref{f7}g and \ref{f7}h, one can see that the preceding sunspot, where the right footpoints of both conjectured MFRs were anchored, has an obvious rotation motion (indicated in the red circles). The rotation angular velocity ($\sim$3$\degr$ h$^{-1}$) results in a maximal flow velocity of $\sim$0.6 km s$^{-1}$ at the edge of the AR where the footpoints of the MFR are located. The clockwise rotation twists the coronal field rooted in this sunspot in the right-handed sense, thus supporting the formation of the MFR. It is however difficult to identify the origin of the rotation, which can be due to either a vortex motion in the photosphere or the emergence of a strongly twisted flux tube, although the probability of the latter is small for a mature AR.

\subsection{Non-Linear Force-Free Field Modeling}\label{ss:NLFFF}
We use an optimization algorithm proposed by \citet{wheatland00} and implemented by \citet{wiegelmann04} to extrapolate the three-dimensional (3D) NLFFF structure of AR 11520. Due to the plasma pressure being dominated by the magnetic pressure in the corona \citep[$\beta \ll$1;][]{gary01}, the coronal magnetic field generally satisfies the force-free criteria, i.e., $\nabla \times \mathbf{B}$=$\alpha \mathbf{B}$ and $\nabla \cdot \mathbf{B}$=0, where $\alpha$ is a constant along each field line. For the magnetic field above ARs, $\alpha$ varies in space \citep{wiegelmann02}. The optimization algorithm minimizes the objective function $L$=$\int_{V} [B^{-2} \vert(\nabla \times \mathbf{B})\times\mathbf{B}\vert^2 + \vert \nabla \cdot \mathbf{B}\vert^{2} ] dV$ through iteration and thus approaches the solution of the NLFFF equations \citep{wiegelmann04}. Before the extrapolation, we apply a preprocessing procedure to the bottom boundary vector data. This removes most of the net force and torque that otherwise generally results in an inconsistency between the forced photospheric magnetic field and the force-free assumption in the NLFFF models \citep{wiegelmann06}.

We compute a time sequence of the 3D NLFFF structure of the AR, covering a period of three days with a cadence of 1 hour. The 3D magnetic field structures at four instants are shown in Figure \ref{f8}a--\ref{f8}h. One can see that at 03:00 UT on July 11, the AR included three sets of field lines: two sets of strongly sheared arcades near the PIL and the overlying constraining field. The right and left arcades correspond very well to the observed double-J sigmoid if seen from above (Figures \ref{f8}a and \ref{f2}a--b). By 07:00 UT on July 12, part of the two groups of arcade field lines may have reconnected, as manifested by continuous sigmoidal field lines (Figure \ref{f8}b), which form a weakly twisted flux rope. At 15:00 UT on July 12, the twist in the rope reached a maximum, since more and more flux was added to the rope by the ongoing reconnection and the rotation of the sunspot (Figure \ref{f8}c and \ref{f8}g). At 12:00 UT on July 13, in spite of the significant decrease of the twist as a consequence of the eruption, the remaining twist still preserved the sigmoidal structure (Figure \ref{f8}d and \ref{f8}h). Overall, the time sequence of 3D NLFFF structures successfully reproduces the evolution of the sigmoid, including the formation and twisting of an MFR before the eruption, as well as the survival of a weakly twisted MFR after the eruption.

\begin{figure*}
\vspace{-0.0\textwidth}
\center {\includegraphics[width=12.5cm]{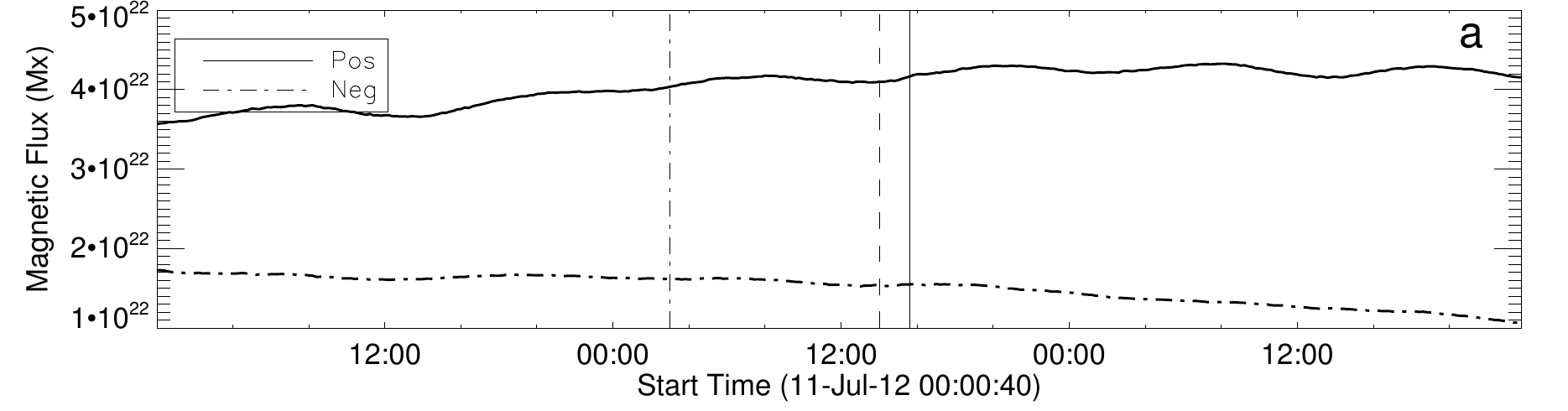}\vspace{-0.02\textwidth}}\\
\center {\includegraphics[width=12cm]{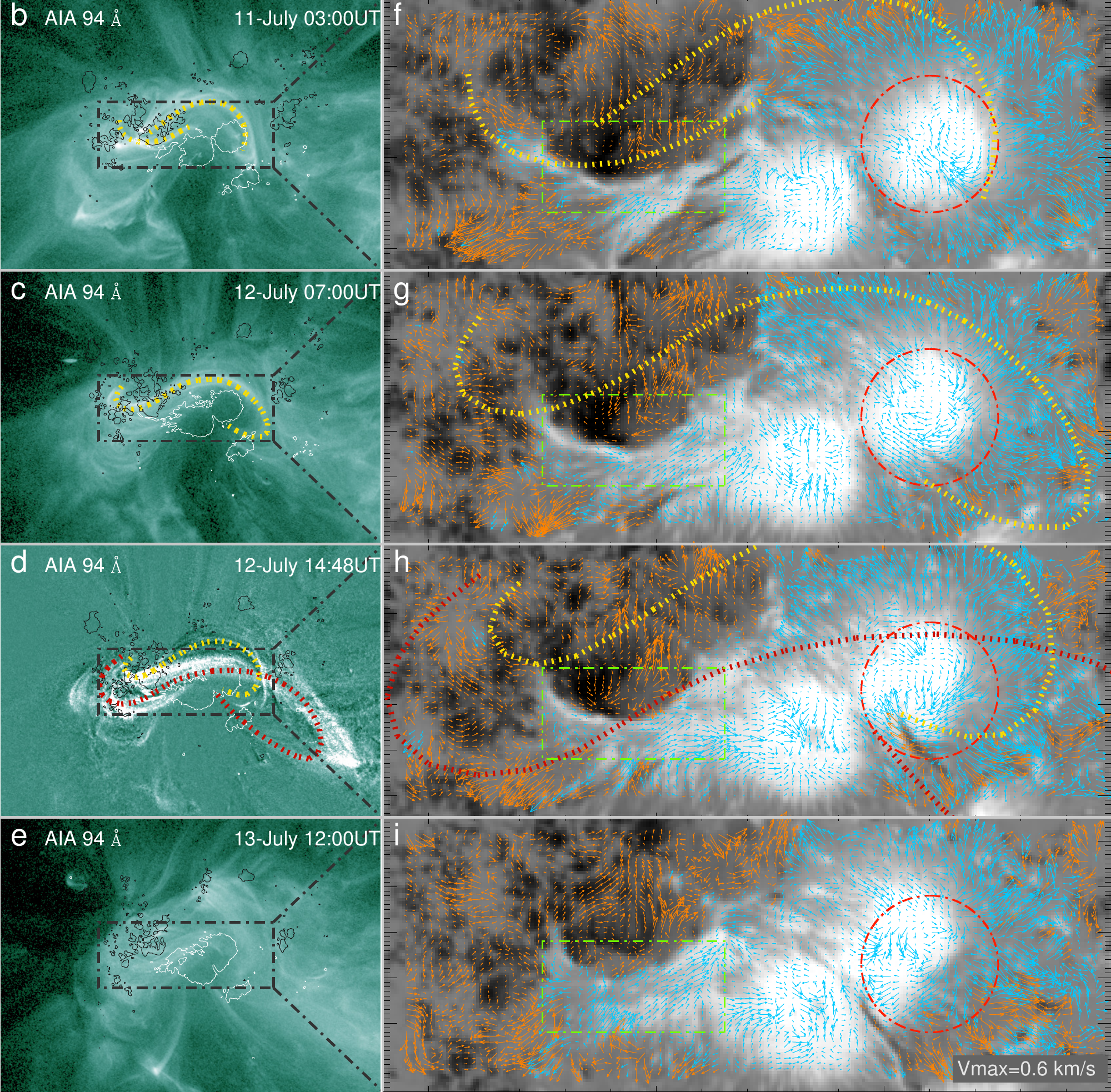}\vspace{-0.0\textwidth}}\\
\caption{a: Temporal evolution of the positive (solid line) and negative (dash-dotted line) magnetic flux in the FOV of panel f--i. The vertical dotted-dash and dash lines show the first appearance of the MFR and the high-lying component of the double-decker MFR. The vertical solid line donotes the onset time of the flare. b--e: AIA 94 {\AA} images overlaid with the contours of the positive (white) and negative (black) polarities of the sunspots. Panel d is an AIA 94 {\AA} base-difference image showing the double-decker MFR close to the eruption. The yellow (red) S-shaped dotted lines indicate the low-lying (the high-lying) MFR. f--i: Line-of-sight magnetograms overlaid by the velocity field at the photosphere and the field lines of the MFR. The green boxes refer to the region with strong shearing flow along the main PIL; the red circles indicate the region with strong rotation.}
(An animation this figure is available in the online journal.)
\label{f7}
\end{figure*}

The NLFFF structures provide the significant clues to the question whether the stable filament resides in a magnetic arcade or in a low-lying MFR. The HMI vector data show that the horizontal field in the section of the PIL under the left half of the sigmoid pointed in the inverse direction (from the negative to the positive side of the PIL). This BP signifies the topology of an MFR that must exist at least in this section of the PIL. The magnetic field modeling suggests that this MFR extended along the whole length of the filament. We conjecture that the sigmoid involved two MFRs, at least in the final two hours before the eruption during which the hot channel was seen. One MFR was high-lying and the other was low-lying (holding the filament); both existed simultaneously above the PIL, constituting a stable double-decker MFR system. From the AIA 94 and 131 {\AA} and XRT images, it is obvious that the two branches of the double-decker system had very closely located footpoints at both ends but rather different lengths (Figure \ref{f2}c).

\begin{figure*}
\vspace{-0.0\textwidth}
\center {\includegraphics[width=14cm]{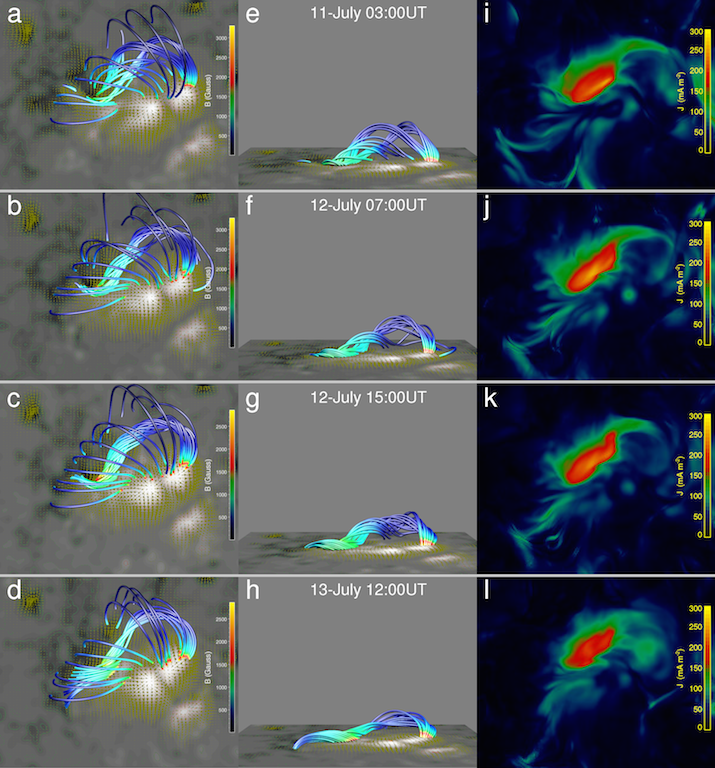}\vspace{-0.0\textwidth}}\\
\caption{Extrapolated 3D NLFFF configurations corresponding to (a) the sheared arcades, (b--c) the pre-eruption sigmoid, and (d) the post-eruption sigmoid. a--d: Top view; e--h: Side view. i-l: Distribution of the current density $|{\bf J}|$ integrated along the line of sight. The bottom boundary is the vertical component of the vector magnetic field overlain by the horizontal component (arrows).}
\label{f8}
\end{figure*}

\begin{figure*}
\vspace{-0.0\textwidth}
\center {\includegraphics[width=14cm]{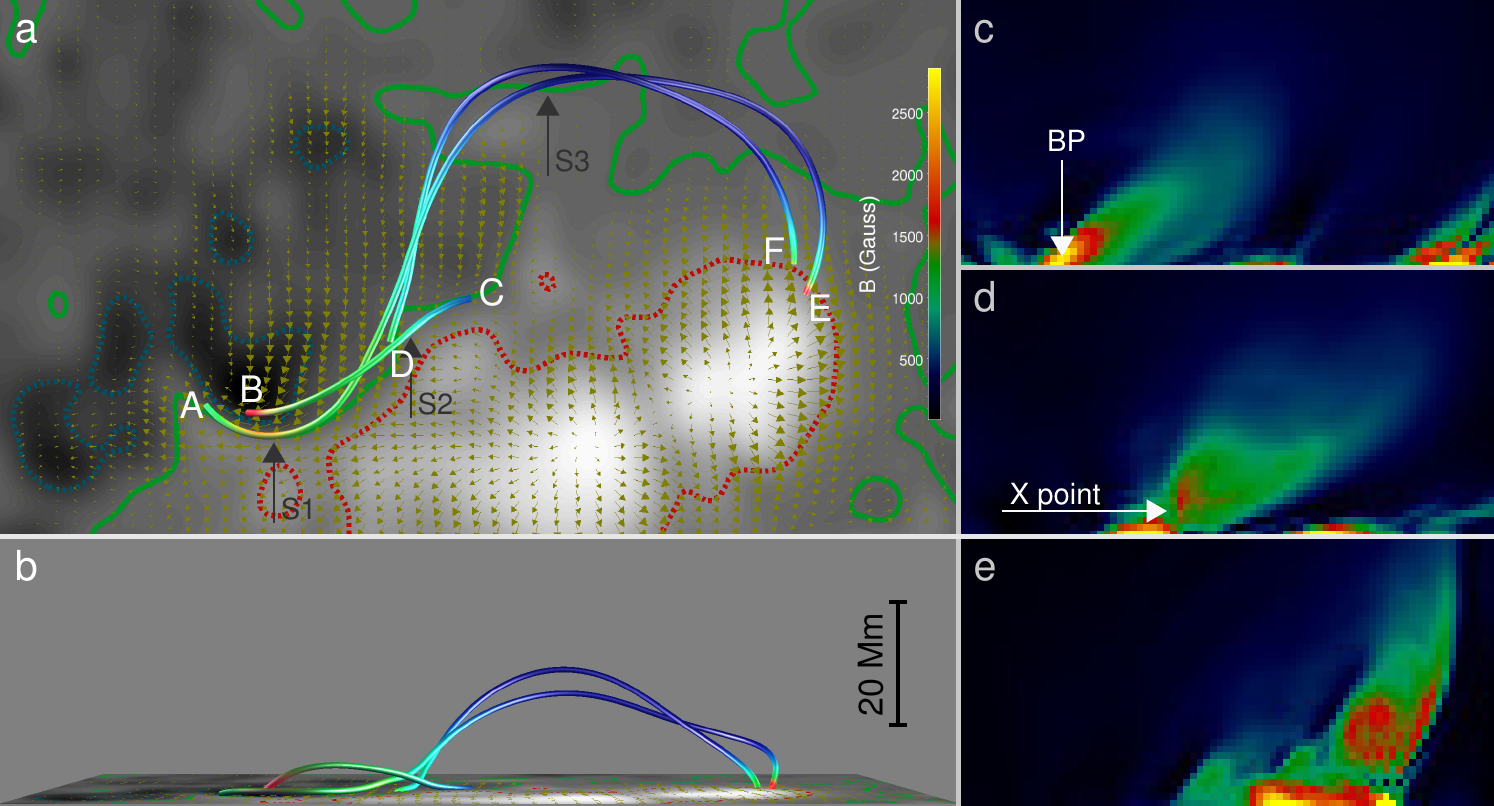}\vspace{-0.0\textwidth}}\\
\caption{a (top view) and b (side view): Representative field lines at 07:00 UT on July 12. The long field line illustrates flux after reconnection, and the two shorter field lines illustrate flux before reconnection. The bottom boundary is the vertical component of the vector magnetic field overlain by the horizontal component (arrows). The blue (red) contours show the magnetic field strength of --1500 G (1500 G), the green contour display the PIL of AR. c--e: Distributions of $|{\bf J}|/|{\bf B}|$ at $x$=S1, S2, and S3 as shown in panel (a). The location of the BP and a conjectured HFT (X point) are indicated in panels (c) and (d), respectively.}
\label{f9}
\end{figure*}

Furthermore, to study the properties of the reconnection during the MFR formation in detail, we plot three representative field lines DF, BC, and AE in Figure \ref{f9}a and \ref{f9}b. DF and BC denote the right and left sheared arcades, respectively. AE shows the S-shaped field of the MFR. To further examine the properties of the reconnection at the different locations, we take three north-south oriented cross sections at $x$=$s_1$, $s_2$, and $s_3$ (Figure \ref{f9}a). The distributions of $|{\bf J}|/|{\bf B}|$ (the total current density normalized by the total magnetic field) at $x$=$s_1$, $s_2$, and $s_3$ are shown in Figure \ref{f9}c--\ref{f9}e, respectively. Here, the current density is given by $\mathbf{J}=\nabla \times\mathbf{B}/\mu_{0}$, where $\mu_{0}$=4$\pi \times$ 10$^{-3}$ G m A$^{-1}$. One can see that $|{\bf J}|/|{\bf B}|$ at the cross section $x$=$s_1$ is mainly concentrated at the BPs (Figure \ref{f9}a--\ref{f9}c), where the horizontal photospheric field components show inverse polarity (Figure \ref{f9}a) and the field lines are concave up with their bottom points touching the photosphere (Figure \ref{f9}b). At the cross section $x$=$s_2$, the distribution of $|{\bf J}|/|{\bf B}|$ displays an X-shape at the height of $\sim$4 Mm, suggesting a hyperbolic flux tube (HFT) in 3D; this indicates that the reconnection mainly takes place in the corona (Figure \ref{f9}d). At the cross section $x$=$s_3$, the location of high coronal current density ascends to a higher position ($\sim$15 Mm) and is mostly concentrated inside the MFR; thus this current has likely a role in heating the MFR. Based on the above properties, we argue that different types of reconnection, i.e., BP, HFT, and internal reconnection, probably exist simultaneously during the formation of the MFR, either generating the twist or heating the plasma. 

We also compare the distribution of the currents with the emission pattern in the sigmoid. Figure \ref{f8}i--\ref{f8}l show the distributions of the current density integrated along the line of sight. It can be seen that the concentration of the currents is mostly along the MFR axis with the largest magnitude appearing at the regions corresponding to the BPs and the HFT. This fact infers that it is the currents that heat up the plasma inside and around the sigmoidal MFR, thus making a sigmoidal emission pattern. Moreover, one can notice that the current density near the main PIL increases before the eruption (Figure \ref{f8}i--j) and decreases afterwards (Figure \ref{f8}k--l). This can be qualitatively explained by the partial eruption of the configuration which released part of the coronal currents.

In addition, we note that the NLFFF model is difficult to reproduce the high-lying MFR in spite of its significant success in reproducing the long-term evolution of the sigmoidal AR. Before the eruption, both the two sets of sheared arcades and the continuous S-shaped and twisted flux bundle are successfully reproduced by the NLFFF model (Figure \ref{f8}a and \ref{f8}b). Even after the eruption, the NLFFF model is still powerful to reconstruct the MFR with a decreased twist that strongly resembles the surviving filament (Figure \ref{f8}d). However, the NLFFF model does not fully succeed in reconstructing the complex magnetic structure near the eruption. The extrapolated 3D structure at 15:00 UT on July 12 only shows a low-lying twisted structure (Figure \ref{f8}c), but misses the high-lying MFR that separated from the low-lying twisted field and appeared as the elongated hot channel structure in the AIA 131 and 94 {\AA} passbands. A possible reason for this partial failure can be ascribed to the proximity of the footpoints of the two twisted coronal structures (Figure \ref{f2}c and Figure \ref{f7}). For a successful reconstruction of both structures, all four footpoint areas must be well resolved by the vector magnetogram, which obviously was not realized.

\section{Summary and Discussion}
In this paper, we investigate the origin of an MFR in the sigmoidal AR that erupted at about 16:10 UT on 2012 July 12 and produced an interplanetary magnetic cloud that caused the strong geomagnetic storm event on July 15. We perform a detailed analysis of the AR's evolution in the three days leading up to the eruption. This includes tracking its morphological evolution, diagnosing the DEM properties, characterizing the motions of magnetic footpoints in the photosphere, and constructing a time sequence of 3D NLFFF models of the coronal structure.

A particularly interesting finding is that two MFRs have been formed above the same PIL of the sigmoidal AR and constituted a stable double-decker MFR system for at least two hours prior to the eruption. The concept of the double-decker MFR was recently proposed by \citet{liurui12_filament} to explain two vertically separated filaments over the same PIL. These authors payed much attention to the identification of the double-decker filament and to the discussion of the mechanisms of the partial eruption of the system. In the present work, we present a second case that supports their new conjecture for the magnetic structure of some CME source regions and additionally concentrate on the formation process of the double-decker MFR. It is found that during a period of the first 40 hours prior to the eruption, an evolving sigmoid manifested the formation of one MFR, most likely the result of reconnection between two groups of sheared arcades near the main PIL. The driver of the reconnection is attributed to the shearing and converging photospheric flows in the vicinity of the PIL, as derived with the DAVE technique. The distribution of the current layers as indicated by the NLFFF extrapolation suggests that the reconnection happens simultaneously at the BPs, i.e., photospheric flux cancellation, and in the HFT in the corona, i.e., tether-cutting. In the present event both worked at the same time in the process of converting the sheared arcades to the twisted field. Besides the shearing and converging flows, the rotation of the leading sunspot probably also played a role in forming the MFR, similar to the vortex motions used for building up twist in some simulations \citep{amari03a,torok03,aulanier05}. A set of continuous S-shaped hot threads indicates that an MFR structure was formed about half a day before the eruption. This low-lying MFR also hosted a filament and remained stable in the eruption.

From about two hours before the eruption we find evidence for the existence of a second MFR in the form of a hot channel \citep{zhang12,cheng13_driver}, which was located above the first MFR. Only the second MFR erupted in the event studied here. The second MFR could be observed by the AIA high-temperature and XRT passbands. It is also imaged in the 195~{\AA} passband, indicating emission in the \ion{Fe}{24} line blend in this channel \citep[$\ge15$ MK;][]{milligan13}. The location above the first MFR excludes its formation by emergence from below the photosphere. The high temperature suggests an important role for reconnection in the formation. This and the proximity of the two MFR in the stable phase prior to the eruption suggest that the double-decker MFR system may have formed by a splitting of the initial and low-lying MFR through the internal reconnection. Such a splitting must be considered as a tentative interpretation, since an MFR is a coherent large-scale structure that generally possesses a considerable degree of stability against perturbations. The low-lying MFR is perhaps most evident from the typically long-lasting stability of quiescent filaments. However, indications that MFR can split or even completely disintegrate do exist, both in observations and numerical modeling. Prominences often show two branches that are clearly separated in height \cite[e.g.,][]{liurui12_filament, suyingna11}. The disintegration of a sigmoid by flux dispersal was described in \citet{tripathi09}. The vertical splitting of an MFR in the evolution to an eruption of only the upper part was found in the numerical modeling of the 1997 May~2 eruption \citep{kliem13}. This evolution was driven by photospheric flows converging at the PIL and enforcing flux cancellation, and it involved slow tether-cutting reconnection with the ambient field in an HFT formed between the splitting parts of the rope. The flux added between the two parts gradually stabilizes (destabilizes) the lower (upper) part of the splitting MFR. \citet{kliem13} further demonstrated that a double-decker MFR system can be in stable equilibrium if the overlying field is sufficiently strong. They also found that the configuration admits a partial eruption, with only the upper branch erupting and the bottom branch remaining stable, very similar to the eruption of the high-lying hot channel and the stability of the low-lying filament. In the present event, strong perturbations of the previously formed MFR in the sigmoid were given by the intrusion of flux along the PIL under the left part of the sigmoid and by the rotation of the sunspot under the right part of the sigmoid, both of which affected the footpoint regions of either MFR. The conjectured internal reconnection must have occurred high enough in the corona \citep[e.g., the second case of Figure 2 in][]{gilbert01} so that the remaining MFR could still support the filament and continuously maintain the sigmoidal pattern of the AR.

The partial eruption of the double-decker MFR system possesses some similarities to but also differences from a partial eruption of a single MFR \citep{gilbert01,gibson06_apjl,gibson08_jgr}. In that case, the reconnection happens in the interface between the MFR and the surrounding fields, e.g., near the crossing point of a kinked MFR \citep{tripathi13}, which results in the escape of the upper part of the MFR with the lower part remaining at the original place. However, one should note that the driver of this internal reconnection is attributed to the helical kink instability, which writhes the MFR axis and generates a current sheet around the MFR \cite[e.g.,][]{kliem10}. In contrast, for the case of the double-decker MFR, photospheric shearing and converging motions drive the reconnection. Therefore, the timescale of the separation is also different; it is nearly instantaneous in the partial eruption model but lasts for hours in the present case.

In the early eruption phase, the morphology of the high-lying MFR varied from an S-shape to a loop-shape, which is very similar to the linear feature in erupting sigmoids \citep[e.g.,][]{moore01,mckenzie08,liur10,aulanier10,green11,zharkov11}. The high similarity suggests that the linear feature is most likely the MFR itself rather than the current shell above the MFR as suggested by \citet{aulanier10}. The difficulty in identifying the linear feature with the MFR in previous studies can be attributed to the unavailability of the high temporal and spatial resolution data.

As the high-lying MFR slowly ascends to a height where the background field declines rapidly enough, the torus instability probably triggers and initiates the impulsive acceleration of the MFR eruption \citep{kliem06,torok05,fan07,aulanier10,olmedo10,savcheva12b,cheng13_double,cheng14_tracking,dudik14}. The initial brightening at the footpoints of the MFR in all EUV and UV passbands is consistent with an enhancement of internal reconnection by the commencing eruption. This is followed by a seamless transition to the much more rapid reconnection in the flare current sheet, in a mutual feedback with the unstable MFR, which not only formed the flare loops further constraining the filament but also produced the high-energy particles that stream down along the newly reconnected loops to generate two well-observed flare ribbons. 

Finally, we find that the NLFFF model of the coronal field, obtained by extrapolation from a sequence of HMI vector magnetograms, succeeds in simulating important aspects in the long-term quasi-static evolution of the sigmoidal AR. The model reproduces the formation of a twisted, sigmoidal flux rope from the highly sheared arcades in very good agreement with the observed coronal structures, and it also resembles the weaker sigmoidal structure after the partial eruption quite well. On the other hand, the double-decker magnetic configuration suggested by the coronal data could not be found. We conjecture that this results from insufficient resolution of the structure in the magnetogram because the footpoints of the two branches were located in close proximity. Moreover, such a sequence of static models may capture complex evolutions only if a much higher time resolution is realized. These facts suggest that more advanced models or MHD simulations should be developed in the future to deal with complex pre-eruption magnetic structures and their dynamic evolution.

\acknowledgements We thank Sarah Gibson, Antonia Savcheva, Yingna Su, David McKenzie, Haisheng Ji, and Yang Liu for valuable discussions, and the referee for constructive comments that helped to improve the manuscript. SDO is a mission of NASAÕs Living With a Star Program. X.C., Y.G., and M.D.D. are supported by NSFC under grants 10933003, 11303016, 11373023, 11203014, and NKBRSF under grants 2011CB811402 and 2014CB744203. X.C. is also supported by Key Laboratory of Solar Activity of National Astronomical Observatories of the Chinese Academy of Sciences by Grant KLSA201311. J.Z. is supported by NSF grants ATM-0748003, AGS-1156120 and AGS-1249270. B.K. acknowledges support by the DFG and by the Chinese Academy of Sciences under Grant 2012T1J0017.


\end{document}